\documentclass{aa}  
\usepackage{graphicx}
\usepackage{txfonts}
\newcommand{\lae}{\mathrel{<\kern-1.0em\lower0.9ex\hbox{$\sim$}}}
\newcommand{\gae}{\mathrel{>\kern-1.0em\lower0.9ex\hbox{$\sim$}}}
 
\begin{document}
   \title{The efficient low-mass Seyfert MCG--05--23--016\thanks{Based on observations with INTEGRAL, an ESA project
   with instruments and science data centre funded by ESA member
   states (especially the PI countries: Denmark, France, Germany,
   Italy, Switzerland, Spain), Czech Republic and Poland, and with the
   participation of Russia and the USA}
 }


   \author{V. Beckmann
          \inst{1,2,3},
           T.J.-L. Courvoisier\inst{1,2},
	   N. Gehrels\inst{4},
           P. Lubi\'nski\inst{5,1},
           J. Malzac\inst{6},
           P.-O. Petrucci\inst{7},
           C. R. Shrader\inst{4},
           \and
           S. Soldi\inst{1,2}
          }

   \offprints{V. Beckmann}

   \institute{ISDC Data Centre for Astrophysics, Chemin d'\'Ecogia 16, 1290 Versoix, Switzerland\\
\email{Volker.Beckmann@unige.ch}
     \and
   Observatoire Astronomique de l'Universit\'e de Gen\`eve, Chemin des Maillettes 51, 1290 Sauverny, Switzerland
     \and
   University of Maryland Baltimore County, 1000 Hilltop Circle,
   Baltimore, MD 21250, USA
     \and
 Astrophysics Science Division, NASA Goddard Space Flight Center, Code
 661, MD 20771, USA
 \and
 Centrum Astronomiczne im. M. Kopernika, Bartycka 18,
PL-00-716 Warszawa, Poland
 \and
   Centre d'\'Etude Spatiale des Rayonnements (CESR), OMP, UPS, CNRS;
   B.P. 44346, 31028 Toulouse Cedex 4, France
   \and
   Laboratoire d'Astrophysique de Grenoble, Univ. J. Fourier/CNRS, UMR
   5571, BP 53X, 38041 Grenoble Cedex, France\\ 
}

   \date{Received: July 25, 2008; accepted: October 16, 2008}

 
  \abstract
   {}
   {The Seyfert 1.9 galaxy MCG--05--23--016 has been shown to exhibit
   a complex X-ray spectrum. 
   This source has moderate X-ray luminosity, hosts a
   comparably low-mass black hole, but accretes at a high Eddington rate,
   and allows us to study a super massive black hole in an early stage.}
   {Three observations of the INTEGRAL satellite simultaneous with
   pointed Swift/XRT observations performed from December 2006 to June
   2007 are used in combination with public
   data from the INTEGRAL archive to study the variability of
   the hard X-ray components and to generate a high-quality spectrum
   from 1 to 150 keV.}
   {The AGN shows little variability in the hard X-ray spectrum, with
   some indication of a variation in the high-energy cut-off energy
   ranging from 50 keV to $\gg 100 \rm \, keV$, with an electron plasma temperature in the $10 - 90 \rm \, keV$ range. The
   reflection component is not evident and, if present, the reflected fraction can be constrained to $R <
   0.3$ for the combined data set. Comparison to previous observations shows that the
   reflection component has to be variable. No variability in the UV and optical range is observed on a
   time scale of 1.5 years.}
   {The hard X-ray spectrum of MCG--05--23--016 appears to be stable
   with the luminosity and underlying power law varying moderately and the optical/UV flux staying constant.  The
   reflection component and the iron K$\alpha$ line 
   seem to have decreased between December 2005 and the observations presented
   here. 
   The spectral energy distribution appears to be similar to that of
   Galactic black hole systems, e.g. XTE~1118+480 in the low state. The AGN
   exhibits a remarkably high Eddington ratio of $L_{bol}/L_{Edd} \gae
   0.8$ (or $L_{bol}/L_{Edd} \gae 0.1$, if we consider a higher mass of
   the central engine) and, at the same
   time, a low cut-off
   energy around 70 keV. Objects like MCG--05--23--016 might indicate the early
stages of super massive black holes, in which a strong accretion flow
feeds the central engine.}

   \authorrunning{Beckmann et al.}
   \titlerunning{The efficient low-mass Seyfert MCG--05--23--016}

   \maketitle


\section{Introduction}

Active galactic nuclei (AGN) are commonly assumed to be super massive
black holes in the centre of galaxies, in which accretion processes
give rise to emission throughout the electromagnetic spectrum. 
AGN are observed to date up to redshifts of $z \sim 6.4$ (Willott et
al. 2007), showing that super massive black holes with masses of $M_{BH} \sim 10^8 \rm \, M_\odot$ must have been formed as early as $< 0.7$
Gyrs after the formation of the first stars (\cite{Kashlinsky05}).
In order to be able to form super massive black holes at this early
stage of the Universe, merging events and high accretion rates are
required. This black hole evolution is closely tied to the growth of the bulge of
the AGN's host
galaxy, as both seem to be correlated with $M_{BH} \simeq
10^{-3} M_{Bulge}$ (\cite{Gebhardt00}).

Recent studies intend to not only find unification models for the
different AGN types but also probe whether their central engines and
super massive black holes are simply up-scaled versions of Galactic
black holes like \object{Cyg X--1}, \object{GRO J1655--40}, or GX 339--4. 
In this context, models of black hole accretion can be tested using
objects showing extreme physical properties, like luminosity,
accretion rate, or Eddington ratio. The Seyfert 1.9 galaxy presented here
resides in an extreme end of the known parameter space of AGN. X-ray
observations give a direct view on matter close
to the super massive black hole, providing insights into the geometry
and the state of the matter. The flux and spectral variability of the sources in the hard X-rays reflect the size and physical state of the regions involved in the emission processes (see Uttley \& McHardy 2004 for a brief review). 

MCG--05--23--016 is one of the brightest Seyfert galaxies in the
X-rays. This Seyfert 1.9 galaxy at redshift $z=0.0085$ has not only
been studied in the $2 - 10 \rm \, keV$ band by most X-ray missions so
far, but has also been detected in hard X-rays above 20 keV with {\it BeppoSAX}/PDS
(e.g. Balestra et al. 2004), {\it INTEGRAL}/IBIS (Soldi et al. 2005), and
{\it Swift}/BAT (Beckmann et al. 2007, Tueller et al. 2008). 
With a bolometric luminosity of $L_{bol} \simeq 2 \times 10^{44} \rm \, erg \,
s^{-1}$ this Seyfert is a moderately luminous object, with a
comparably small central black hole of $M_{BH} = 2 \times 10^6 \rm \,
M_\odot$ (\cite{BHmass}). Wang \& Zhang used the
width of the OIII line (\cite{Greene05}) to obtain the black hole
mass from the $M_{BH} - \sigma$ relation (\cite{Tremaine02}) and
give the error of the measurement with 0.7~dex. 

Balestra et al. (2004) derived from {\it ASCA}, {\it BeppoSAX}, {\it
  Chandra} and {\it XMM-Newton} data that the iron K$\alpha$ line
apparent in the spectrum seems to be a superposition of a narrow and a
broad component, that a reflection component of $R = 0.45$ is
detectable, and that the continuum can be modeled by a photon index of
$\Gamma \simeq 1.7$ and a cut-off at $E_C = 110 \rm \, keV$. The
  observations covered a time span from 1994 to 2001, during which the object showed little flux variation and a
stable spectrum as well as constant iron line and reflection
  component. A recent re-analysis of the {\it BeppoSAX} data by Dadina
  (2007) indicates
  a slightly steeper spectrum ($\Gamma = 1.79 {+0.07 \atop -0.08}$)
  with higher cut-off energy ($Ec = 191 {+110 \atop -60} \rm \, keV$), and
  stronger reflection component ($R = 0.74 {+0.22 \atop -0.52}$). An
  analysis of {\it RXTE} data by Mattson \& Weaver (2004) using a fixed underlying power law slope
  of $\Gamma = 1.88$ led to a variable reflection component in the
  range $R = 0.35 - 0.57$ and iron $K\alpha$ line flux ($f_{K\alpha} =
  (1.55 - 1.99) \times 10^{-4} \rm \, photons \, keV^{-1} \, cm^{-2}
  \, s^{-1}$).
Recently, observations on MCG--05--23--016 by {\it Suzaku} showed a
  $0.4-100 \rm \, keV$ spectrum which appeared to be a cut-off power
  law of $\Gamma = 1.9$ and $E_C > 170 \rm \, keV$ plus a dual
  reflector with $R \sim 0.9$ and $R \sim 0.5$, respectively
  (\cite{Reeves07}). {\it XMM-Newton} and {\it Chandra} data confirmed that
  the iron K$\alpha$ line complex consists of a broad and of a narrow
  component and revealed evidence for outflowing material
  (\cite{Braito07}).
The exact shape and structure of the high-energy spectrum of
  MCG--05--23--016 remains elusive though. {\it INTEGRAL} and {\it
  Swift} observations
  can help to shed light on the existence and position of the
  high-energy cut-off and variability of the hard X-ray spectrum.
  In Section~2 we describe the analysis of the {\it INTEGRAL} and
  simultaneous {\it Swift} observations. We discuss the findings in Section~3,
  with emphasis on similarities with Galactic black hole systems and
  on the fact that the Seyfert galaxy apparently
  operates at high Eddington ratio. Conclusions are presented in Section~4.


\section{Observations and data analysis}

The {\it INTEGRAL} (\cite{INTEGRAL}) mission offers the unique
  possibility to perform simultaneous observations of AGN over the $3 - 1000
  \rm \, keV$ energy region. This is achieved by the X-ray monitor
  (3--35 keV) JEM-X (\cite{JEMX}), the soft gamma-ray imager (18--1000
  keV) IBIS/ISGRI (\cite{ISGRI}), and the spectrograph SPI (\cite{SPI}),
  which operates in the 20 -- 8000 keV region. Each of these
  instruments employs the coded-aperture technique. 

{\it INTEGRAL} performed three dedicated pointed observations on
\object{MCG--05--23--016} in December 2006 (spacecraft revolution 508) and in
May/June 2007 (revolution 565 and 569). As observing strategy
hexagonal dithering has been chosen in order to achieve a sufficient
coverage also by the JEM-X monitor. During these observations the
spectrometer SPI was not available because it went through an annealing
cycle. 
In addition and for comparison reasons, we analysed {\it INTEGRAL}
data on MCG--05--23--016 which had been taken prior to our
observations. During all these measurements the source was too far
off-axis to be covered by JEM-X. 
Reliable data from the SPI
spectrometer can be taken also at larger off-axis angles. Therefore
the public SPI data include more revolutions than the IBIS/ISGRI data
analysed here.
All {\it INTEGRAL} data were analysed using the Offline Science Analysis (OSA)
software version 7.0 distributed by the ISDC (\cite{ISDC}). 
For IBIS/ISGRI standard spectral extraction was applied, using 12
spectral bins in
the 18 keV to 521 keV energy range. For
JEM-X the flux values were extracted from imaging analysis in four
energy bands (3 -- 6 -- 10 -- 15 -- 35 keV), which
provides more reliable results for a source at the flux level of
MCG--05--23--016 than standard spectral extraction.

In order to achieve a wider spectral energy coverage, pointed {\it Swift}
observations were performed simultaneous with the {\it INTEGRAL}
ones. {\it Swift} (\cite{Swift}) provides two narrow-field instruments,
the XRT (\cite{XRT}) with
energy range 0.3--10.0 keV and the UVOT (\cite{UVOT}) which is equipped with 6
filters in the optical and UV range. The XRT and UVOT data have been
analysed using HEADAS version 6.1.1 applying latest calibration files
as available in January 2008. 

\begin{table*}
\caption[]{Observation log of {\it INTEGRAL} and {\it Swift}/XRT observations on
  MCG--05--23--016}
\label{obslog}
\begin{tabular}{ccrrcr}
{\it INTEGRAL}   & time covered  & ISGRI exposure & JEM-X exp. & Swift/XRT  & exposure\\
revolution & U.T.          &  [ks]    &    [ks]    &  U.T.      &  [ks]\\
\hline
256 -- 282 & 04-11-18 -- 05-02-05 &  69.3 &   --  & 05-12-09 & 10.3 \\
508 & 06-12-12 -- 06-12-13        &  26.3 &  21.7 & 06-12-13 & 1.9  \\
565 & 07-05-30 -- 07-06-01        & 128.7 & 122.0 & 07-05-31 & 2.1  \\
569 & 07-06-12 -- 07-06-13        &  89.4 &  79.8 & 07-06-13 & 2.9  \\
\end{tabular}
\end{table*}

The {\it INTEGRAL} IBIS/ISGRI and JEM-X1, as well as the {\it Swift}
pointed observations are summarised in Table~\ref{obslog}. The
observations in {\it INTEGRAL} revolution 508, 565, and 569 are truly
simultaneous with {\it Swift} pointed observations,
while the public data obtained from revolution 256 to 282 are separated
from the {\it Swift} observation by at least 10 months and also do not
include any JEM-X data and therefore no information about the 9--18 keV
energy range. 
The SPI data showed only a marginal detection on the $4\sigma$ level
in the 20--40 keV energy band in the combined data set. The SPI
spectrum can be fit by a single power law model with photon index
$\Gamma = 1.8 {+1.5 \atop -0.6}$\footnote{errors are $2\sigma$ errors
  throughout the paper.}. As these data
are not strictly simultaneous with the other observations and are of
low significance, the SPI data have not been considered in the following
analysis of MCG--05--23--016.

\begin{figure}
\centering
\includegraphics[height=8.5cm,angle=-90]{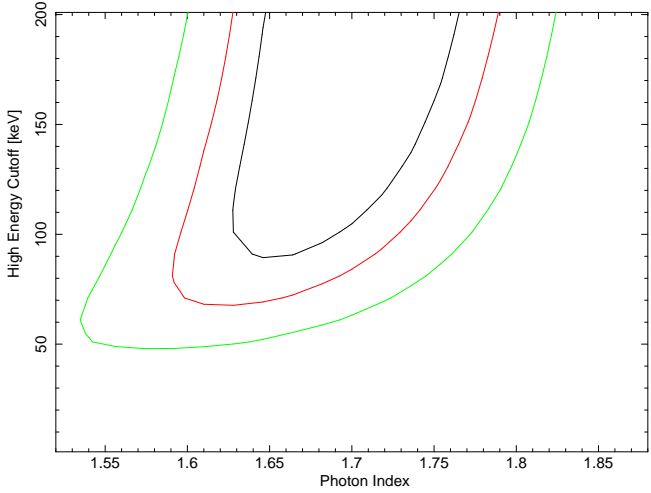}
\caption{Contour plot for an absorbed cut-off power law applied to the
  combined {\it Swift}/XRT, {\it INTEGRAL} JEM-X1, and
  IBIS/ISGRI data obtained in December 2006. The contours show the
  allowed range for the photon index and the cut-off energy at the 1,
  2 and 3 sigma level. No
  cut-off is required by the data.}
              \label{rev508_contourplot}
\end{figure}

The data of IBIS/ISGRI, JEM-X, and Swift/XRT have been fit
simultaneously using XSPEC version 11.3.2 (\cite{XSPEC}). As the most simplistic
approach an absorbed power law model has been fit to the data. Only in
the case of revolution 508 data this model gives an acceptable fit with $\chi_\nu^2 = 1.06$ for 55 degrees of freedom. In the
other cases, a good fit was achieved with an absorbed cut-off power
law model. 
The fit results are summarised in Table~\ref{fitresults}.
\begin{table}
\caption[]{Spectral fitting of combined {\it Swift}/XRT and INTEGRAL
  ISGRI/JEM-X1 data of MCG--05--23--016 using an absorbed cut-off power
  law model}
\label{fitresults}
\begin{tabular}{cccccc}
{\it INTEGRAL}   & $N_{\rm H}$  & $\Gamma$ & $E_C$ & $\chi^2_\nu$ & d.o.f. \\
revolution & $[10^{22} \, \rm cm^{-2}]$ & & [keV] & & \\
\hline
256 -- 282 & $1.26 {+0.07 \atop -0.07}$ & $1.49 {+0.08 \atop -0.08}$ &
$57 {+31 \atop -16}$ & 1.05 & 276\\
508 & $1.29 {+0.15 \atop -0.14}$ & $1.76 {+0.08 \atop -0.08}$ & $>90 $ &
1.06 & 55\\
565 & $1.24 {+0.14 \atop -0.13}$ & $1.51 {+0.09 \atop -0.10}$ & $86 {+68 \atop -29}$ & 0.95 & 80\\
569 & $1.04 {+0.10 \atop -0.09}$ & $1.42 {+0.09 \atop -0.09}$ & $65
{+35 \atop -18}$ & 0.92 & 97\\
all data & $1.24 {+0.05 \atop -0.05}$ & $1.52 {+0.07 \atop -0.05}$ & $72
{+21 \atop -14}$ & 0.99 & 268\\
\end{tabular}
\end{table}
All spectra require a high-energy cut-off in the
range 60--85 keV. The only exception is the short exposure in revolution
508. In this case, the existence of a cut-off at 50 keV, 70 keV and 90
keV, can be rejected at a
significance level of $3\sigma$, $2\sigma$, and $1\sigma$,
respectively, as shown in Figure~\ref{rev508_contourplot}.
None of the spectra required an additional reflection
component. Applying a Compton reflection model, such as {\tt pexrav}
(\cite{pexrav}) which describes a reflection component from cold
material using a cut-off power-law as continuum,
shows that the existence of such a component in MCG--05--23--016 can neither be proven nor
rejected by the single spectra. The soft X-ray excess below $0.8 \rm
\, keV$, as reported in Reeves et al. (2007), is apparent only in {\it
  Swift}/XRT spectrum of December 2005.
 
\begin{figure}
\centering
\includegraphics[height=8.5cm,angle=-90]{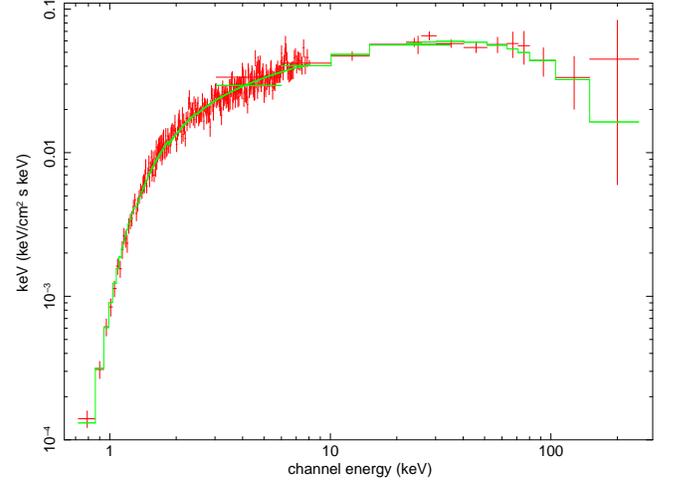}
\caption{Combined spectrum of all {\it Swift}/XRT, {\it INTEGRAL} JEM-X1, and
  IBIS/ISGRI data obtained between November 2004 and June 2007. The
  absorbed cut-off power law fit results in $N_{\rm H} = (1.24 \pm
  0.05) \rm \, 10^{22} \rm \,
cm^{-2}$, photon index $\Gamma = 1.52 \pm 0.05$, and cut-off
energy $E_c = 72 {+21 \atop -14} \rm \, keV$ for a $\chi^2_{268} = 0.99$.}
              \label{fig:allcombined}
\end{figure}

\begin{figure}
\centering
\includegraphics[height=8.5cm,angle=-90]{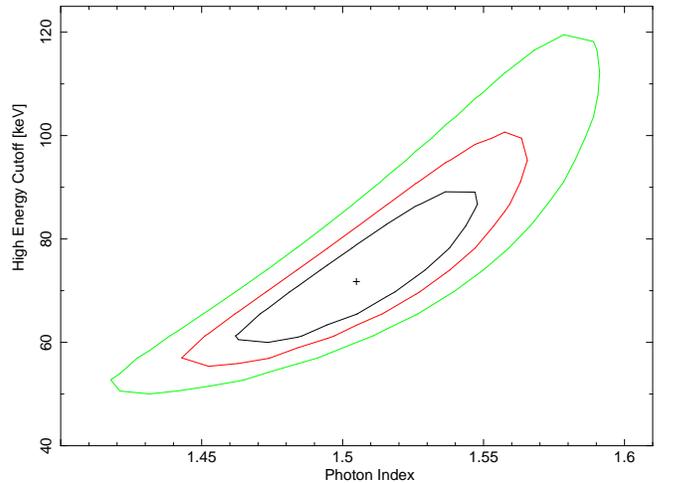}
\caption{Contour plot for an absorbed cut-off power law applied to all
  the
  combined {\it Swift}/XRT, {\it INTEGRAL} JEM-X1, and
  IBIS/ISGRI data listed in Table~\ref{obslog}. The contours show the
  allowed range for the photon index and the cut-off energy at the 1,
  2 and 3 sigma level.}
              \label{fig:alldata_contourplot}
\end{figure}

As the spectral parameters listed in Tab.~\ref{fitresults} appear to
be nearly constant, we also applied a fit to the total combined data
set as shown in Figure~\ref{fig:allcombined}. The total combined spectrum can be fit by an absorbed cut-off power law
model with $N_{\rm H} = (1.24 \pm 0.05) \rm \, 10^{22} \rm \,
cm^{-2}$, photon index $\Gamma = (1.52 \pm 0.05)$, cut-off
energy $E_c = 72 {+21 \atop -14} \rm \, keV$. The fit results in
$\chi^2_{268} = 0.99$. The significance of the high energy cut-off and
the tight constraint on the spectral slope can be seen in the contour
plot shown in Figure~\ref{fig:alldata_contourplot}.
For this combined spectrum also the strength of a
possible Compton reflection component can be constrained. 
Applying the {\tt pexrav} model and using the
same relative iron abundance of $0.4$ and inclination angle of
$45^\circ$ as Reeves et al. (2007) result in similar absorption, photon index and
cut-off energy ($N_{\rm H} = 1.25  {+0.06 \atop -0.05} \rm \, 10^{22} \rm \,
cm^{-2}$, $\Gamma = 1.52 {+0.07 \atop -0.05}$, $E_f = 72 {+21 \atop
  -14} \rm \, keV$), and gives a relative reflection strength of $R =
0.06 {+0.20 \atop -0.06}$ ($\chi^2_{267}= 0.99$). 
The {\tt compPS} model developed by Poutanen \&
Svensson (1996) for a  plane-parallel slab geometry using exact numerical
solution of the radiative transfer equation indicates a stronger reflection component ($R = 0.4
{+0.3 \atop -0.2}$ with an electron temperature of $kT_e = 113
{+15 \atop -10} \rm \, keV$ ($\chi^2_{265}= 1.04$). It has to be pointed out though that this model
is a significantly worse representation of the data than the {\tt
  pexrav} or the simple cut-off power law model.
Thus, also in the
overall combined spectrum a reflection component is not required, and
can be constrained to $R < 0.25$ on a $2\sigma$ level. 
Adding a Gaussian line at the position of the FeK$_\alpha$ line gives
an equivalent width of $EW = 61 {+47 \atop -40} \rm \, eV$, although
this component is not statistical significant.

We then left the iron abundance as a free parameter in the {\tt
    pexrav} model and forced the iron
line flux to have an equivalent width of 100 eV with respect to the
reflection continuum alone. Applying this model, one finds that the
    iron abundance is not constrained by the data and one gets
similar results with a marginal evidence for a reflection component of
$R = 0.13 {+0.18 \atop -0.13}$ consistent with no reflection
, a cut-off energy of $E_f = 77 {+17 \atop
  -18} \rm \, keV$, and photon index $\Gamma = 1.56 {+0.05 \atop -0.09}$ ($\chi^2_{265}= 1.02$).

 
The {\it Swift} pointed observations also provide flux information about
the $UV$ and the optical $V$, $B$, and $U$ bands. All four {\it Swift}/UVOT
measurements revealed the same fluxes within the error of $0.1 \rm \,
mag$. The results were: $V = 13.9 \rm \, mag$, $B = 14.9 \rm \, mag$,
$U = 15.3 \rm \, mag$, $UVW1 = 16.1 \rm \, mag$, $UVM2 = 17.0 \rm \,
mag$, and $UVW2 = 16.9 \rm \, mag$.

\subsection{Detection of 1RXS J092418.0--314212}

Within the JEM-X data set covering the 3--35 keV energy range, another
source is clearly detected in the field of view during the observation
of MCG--05--23--016. The source position
coincides with \object{1RXS J092418.0--314212}, currently an unidentified
object. The source is not detectable in the IBIS/ISGRI data
set and the source position is not within the {\it Swift}/XRT field of view. 
The JEM-X data cannot be represented by a power law model, but are
well represented by a blackbody model with temperature $kT = 1.4 {+0.3
  \atop -0.2} \rm \, keV$. We therefore suggest that this source is
likely to be a Galactic X-ray binary system and that the spectrum
shows the black body radiation emitted from the central parts of
its accretion disc.

\section{Discussion}

\subsection{The hard X-ray emission}

In our study of MCG--05--23--016 we have made use of the hard X-ray
capabilities of {\it INTEGRAL}, and in particular of the imager
IBIS/ISGRI, in combination with the soft X-ray spectrum provided by
{\it Swift}/XRT. Although the combined spectrum clearly does not
provide the wealth of information in the soft X-rays as shown in {\it
  XMM-Newton}, {\it Chandra}, and {\it Suzaku} data, the better
sensitivity at energies above 30 keV puts significant constraints
on the hard X-ray component. 
First of all, a high-energy cut-off in the range 57 keV to 86 keV is required in most of the
observed spectra. However, one observation in December 2006 shows a
spectrum extending up to $\sim 200 \rm \, keV$, excluding 
a cut-off at 50 keV on a $3\sigma$ level, and at 90 keV on a $1\sigma$
level, even when allowing for a different photon index of the
underlying power law at the same time
(Fig.~\ref{rev508_contourplot}). It is important to note though, that
the spectrum in this case is significantly steeper ($\Gamma = 1.7 \pm
0.1$) than in the cases
where a cut-off is measured ($\Gamma = 1.5 \pm
0.1$). 

The differences in the hard X-ray spectrum, i.e. the differences in
high-energy cut-off and spectral slope, can
be interpreted as different temperatures of the electron plasma, which
is the source of the inverse Compton emission thought to be the
dominant component in this energy range.
A physical model describing 
Comptonization of soft photons by a hot plasma, the so called {\tt
  compTT} model, has been developed by Titarchuk (1994). The model
includes the plasma temperature $T_e$ of the hot corona, the optical
depth $\tau_p$ of this plasma, and the temperature $T_0$ of the soft
photon spectrum. Because the spectrum starts at $E \simeq 0.7 \rm \, keV$,
$T_0$ is not well constrained by the data and has been fixed to 10 eV.
Table~\ref{compTTresults} gives the results of this model for the
individual data sets as well as for the combined spectrum  for plane geometry.
\begin{table}
\caption[]{Thermal Comptonisation model (compTT) applied to combined MCG-05-23-016 data}
\label{compTTresults}
\begin{tabular}{ccccccc}
{\it INTEGRAL}   & $N_{\rm H}$  & $kT_e$ & $\tau_p$ & $L_{2-100 \rm \, keV}$ & $\chi^2_\nu$ & d.o.f. \\
revolution & $[10^{22} \, \rm cm^{-2}]$ & [keV] & & $[\rm erg \,
s^{-1}]$ & & \\
\hline
256 -- 282 & $1.33 {+0.07 \atop -0.07}$ & $11 {+15 \atop -3}$ & $3.4
{+0.8 \atop -1.0}$ & $4.3 \times 10^{43}$ & 1.08 & 275\\
508 & $1.21 {+0.13 \atop -0.11}$ & $90 {+76  \atop -73}$& $0.5
{+2.4 \atop -0.4}$ & $4.6 \times 10^{43}$  & 0.97 & 49\\
565 & $1.35 {+0.11 \atop -0.11}$ & $23 {+105 \atop -7}$ & $2.1 {+0.6 \atop
-1.8}$ & $4.2 \times 10^{43}$& 0.91 & 79\\
569 & $1.14 {+0.08 \atop -0.08}$ & $14 {+5 \atop -3}$ & $3.1 {+0.6 \atop
 -0.5}$ & $4.4 \times 10^{43}$& 0.92 & 96\\
all data & $1.33 {+0.04 \atop -0.04}$ & $18 {+7 \atop -3}$ & $2.5
{+0.4 \atop -0.6}$ & $4.4 \times 10^{43}$& 1.04 & 268\\ 
\end{tabular}
\end{table}
As for the cut-off energy $E_C$, the variation of the plasma
temperature is significant on a $2\sigma$ level. The highest plasma
temperature and lowest optical depth is observed during revolution 508, when the cut-off in the
hard X-ray spectrum disappeared (Tab.~\ref{fitresults}). However, it
has to be taken into account that the rev. 508 spectrum can also be fit by
applying the same {\tt compTT} model which gives the best fit to all data (i.e. with $kT_e
= 18 \rm \, keV$ and $\tau_p = 2.5$) with an acceptable fit result
($\chi^2_\nu = 1.09$ for 52 d.o.f.).  

Similar values for cut-off energy and photon index as presented here were derived
previously based on combined {\it BeppoSAX} and {\it INTEGRAL}
IBIS/ISGRI data, resulting in $E_C = 112 {+40 \atop -43} \rm \, keV$,
$\Gamma = 1.74 {+0.08 \atop -0.14}$, plus a reflection component with
$R = 1.2 {+0.6 \atop -1.0}$ (\cite{Molina06}). As these observations
are not simultaneous, these values have to be taken with some caution
though, and the reflection component is not well constrained. 

\begin{figure}
\centering
\includegraphics[height=8.5cm,angle=-90]{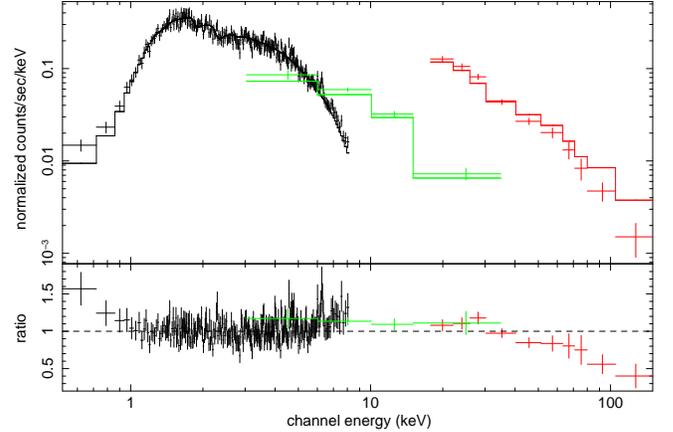}
\caption{Absorbed single power law model fit applied to the $2-5 \rm
  \, keV$
  range of the combined
  data set. The ratio between the data and the model shows the soft
excess at energies below 1 keV, the iron K$\alpha$ line, the
significant cut-off at high energies, but no trace of a reflection component.}
              \label{fig:ratioplot}
\end{figure}

Apart from the variation of the cut-off energy and of the plasma
temperature, describing a variable inverse Compton process, the
  low level of the reflection component in the data studied here is remarkable. 
Figure~\ref{fig:ratioplot} shows a simple model fit applied to the
combined data set, using an absorbed power law model fitted in the 2 -
5 keV range as done by Reeves et al. (2007) to show the significance
of the reflection component. The ratio
between the data and the model shows several components: the soft
excess at energies below 1 keV, the iron K$\alpha$ line, and the
significant cut-off at high energies. There is no evidence however of
a significant reflection component. 
The $2\sigma$ upper limit is $R < 0.25$, while in the observations
presented by Reeves et al. (2007), a value of $R< 0.7$ can be excluded
on a $3\sigma$ level. 
We have to conclude that the strong reflection component, visible in
the December 2005 data taken by {\it Suzaku}, {\it XMM-Newton} and
{\it Chandra} disappeared within a year. Also the excess observed at
soft X-rays by Reeves et al. is only detectable in the {\it Swift}/XRT data from
December 2005 and not afterwards. At the same time, also the
iron line complex has decreased in flux, from an equivalent
width of $EW = 130 \pm 17 \rm \, eV$ in December 2005 to about $EW =
61 \pm 45 \rm \, eV$ in 2006/2007. Unfortunately the {\it Swift}/XRT
data do not allow us to disentangle the broad and narrow line component
of the line complex, as reported in Balestra et al. (2004) and also in
Reeves et al. (2007). If we assume that the narrow and the broad line
components arise from two different absorber, with the broad line
being emitted closer and the narrow line further away from the central
engine as proposed by Reeves et al., then we might indeed
observe the disappearance of the broad iron line ($EW \simeq 60 \rm \,
eV$; Reeves et al. 2007) and of the reflection component connected to
the close absorber between the December 2005 and December 2006 observations,
leaving the spectrum only with the narrow line component and
reflection from the distant absorber.
This disappearance is not accompanied by a significant change
in X-ray luminosity or in UV/optical flux. The lack of hard X-ray flux 
variability appeared also in a study of MCG--05--23--016 using {\it
  Swift}/BAT, which resulted in a marginal variability of $6 \pm 4 \%$
on a 20 day time scale (\cite{BATvariability}).
The soft X-ray spectrum in the range 2--10 keV however
appears to be variable by a factor of $\sim 1.7$ throughout historic observations. The {\it Swift}/XRT data
presented here give a flux of $f_{2-10 \rm \, keV} = (7.1 - 8.1)
\times 10^{-11} \rm \,
erg \, cm^{-2} \, s^{-1}$, with literature values in the range
$f_{2-10 \rm \, keV} = (7.1 - 11.9) \times 10^{-11} \rm \, erg \,
cm^{-2} \, s^{-1}$ (e.g. Mattson \& Weaver 2004, Balestra et al. 2004). 

Mattson \& Weaver (2004) also found variable Compton reflection when
studying MCG--05--23--016 with {\it RXTE} and {\it BeppoSAX}. In their
study, the strength of the reflection component is anti-correlated
with the $2 - 10 \rm \, keV$ flux. In this context, one would expect a
very high reflection component in the data presented here, where the
2--10 keV flux is at the historic minimum. 
When using
the same photon index of $\Gamma = 1.77$ as Mattson \& Weaver, we get
a relative reflection of $R = 0.78 {+0.26 \atop -0.21}$, but at the
same time $\chi^2_\nu$ rises to 1.10, compared to $\chi^2_\nu = 0.99$
with no reflection and a photon index of $\Gamma = 1.52$. We therefore
do not consider the fixed spectral slope the valid approach to fit the
data presented here. 
A similar behaviour, of a hardening of the spectra with decreasing
  reflection component as we see it from comparison with previous measurements, has been observed before. Zdziarski, Lubinski
  \& Smith (1998) analysed this relation for 61 {\it Ginga}
  observations of 24 radio quiet Seyfert galaxies. They found a strong
  correlation of the form $R = u \, \Gamma^v$ with the parameters $u =
  (1.4 \pm 1.2) \, 10^{-4}$ and $v = 12.4 \pm 1.2$. The observed trend
  follows closely this relation. Our measurement of $\Gamma = 1.5$
  would result in $R= 0.03 \pm 0.04$, consistent with the upper limit
  we derived. Also the results of Mattson \& Weaver and of Reeves et
  al. (2007) fit well on this correlation. The measurement of Molina
  et al. (2006) based on non-simultaneous {\it BeppoSAX} and {\it INTEGRAL}
IBIS/ISGRI data however, with $R = 1.2$ and $\Gamma = 1.7$ lies about
$7 \sigma$ off this relation which results in $R = 0.13 \pm
0.16$. This might indicate that it is indeed essential to study
simultaneous observations when determining the strength of the
reflection component.

An inverted source behaviour in the sense that the strength of the reflection component is anti-correlated
with the $2 - 10 \rm \, keV$ flux might be another hint for a
scenario in which the reprocessing occurs at large distance to the
central engine of the AGN, and therefore a correlation of the
reflected component and the continuum would appear with a large delay
(\cite{Malzac02}). Based on the data analysed here, this distance
appears to be 1.5 light years, as there is no variability
observable in the 2--100 keV X-ray luminosity, but significant
variation in the reflection component. This would place the material
relevant for the reprocessing at a distance of $0.55 \rm \, pc$, at the outer regions of the accretion disc. This being said, we cannot
exclude that the flux varied in between the observations presented
here, therefore the distance has to be taken as an upper limit.

\begin{figure}
\centering
\includegraphics[height=8.5cm,angle=0]{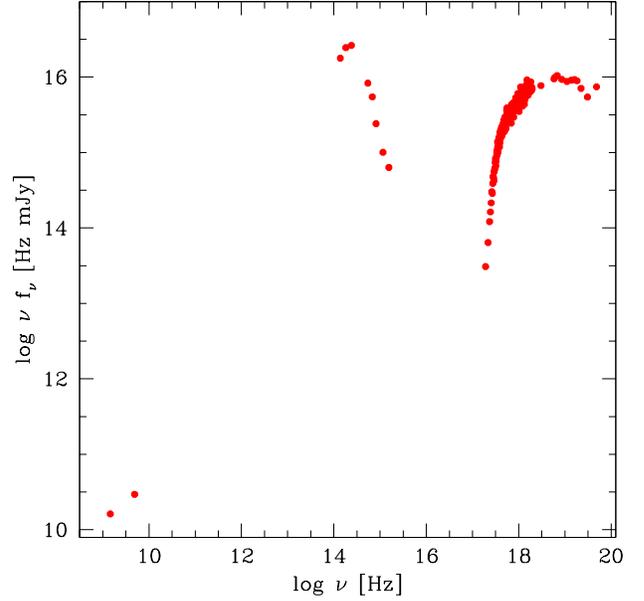}
\caption{Spectral energy distribution of the combined data set,
  including {\it Swift}/UVOT optical and UV data as well as {\it
  Swift}/XRT soft X-ray, and {\it INTEGRAL} JEM-X and IBIS/ISGRI hard
  X-ray data. In addition, 2MASS measurements (J, H, and Ks) and
  radio data (6 cm and 20 cm) have been added. Data have been corrected for Galactic hydrogen column density, but
  not for intrinsic absorption.}
              \label{fig:SED}
\end{figure}

\subsection{Accretion power in MCG--05--23--016}

MCG--05--23--016 presents a rather unique case. Its spectrum is, at
least to the first order, nearly constant in shape and luminosity,
with a possibly variable reflection component ($R = 0 \dots 1$) and
high energy cut-off ($E_C = 50 \dots 120 \rm \, keV$). The over all
luminosity is low. Using the X-ray luminosity of $L_{2-200 \rm \, keV} =
10^{44} \rm \, erg \, s^{-1}$ as a proxy for the bolometric luminosity, the small black hole mass of $2 \times 10^6 \rm \,
M_\odot$ (Whang \& Zhang 2007) leads to a large Eddington ratio of
$L_{bol}/L_{Edd} \gae 0.4$. As the UV/optical emission is at least of the same
order as the X-ray one, it is more likely that the Eddington ratio is
as high as $L_{bol}/L_{Edd} \gae 0.8$. Considering the error on the mass
determination, this translates in a range of $L_{bol}/L_{Edd} = 0.2 -
5$. 

An independent way to determine the mass of a black hole is to
  study its temporal behaviour. Following McHardy et al. (2006),
  the detection of a break in the X-ray power spectral density (PSD)
  of an AGN would allow to estimate the black hole mass thanks to
  the relation $T_B = 2.1 \log (M_{BH} \times 10^{-6} M_\odot^{-1}) - 0.98 
  \log (L_{bol} \times 10^{-44} \rm \, erg^{-1} \, s) - 2.32$ .
  Using the data collected by {\it RXTE}/PCA and {\it XMM-Newton}/PN between
  1996 and 2005, we estimated the X-ray PSD and structure function ({\it RXTE} data only)
  of MCG--05--23--016.
  The structure function has the advantage of working in the time domain and therefore being less
  sensitive to alias and windowing problems than the Fourier analysis.
  The PSD was calculated for each observation longer than 10 ks (on 200 seconds binned light curves),
  and the PSDs obtained were averaged and binned in logarithmically spaced bins (e.g. Uttley et al. 2002).
  The final PSD shows a flattening between $2 \times 10^{-5} - 10^{-4} \, \rm Hz$ (10--50 ks), 
  close to the minimum frequencies sampled by these data ($\sim 7 \times 10^{-6} \, \rm Hz$).
  Also the structure function presents the hint of a break, but rather at $\sim$3.5 ks.
  The lack of a longer-term monitoring prevents us from drawing 
  a firm conclusion about the presence and the position of the break.
  Assuming a break in the range 0.04--0.6 days (3.5--50 ks), we achieve 
  a black hole mass of $M_{BH} = 1.84-3.3 \times 10^7 \, \rm M_\odot$.
  It has to be taken into account though that for the given bolometric
  luminosity the formula does not allow for black hole masses
  lower than $1.8 \times 10^7 \, \rm M_\odot$, even when the break
  time approaches zero. This means that it implies an upper
  limit for the Eddington ratio of $L_{bol}/L_{Edd} \lae 0.1$ at a
  luminosity of $L_{bol} = 2 \times 10^{44} \rm \, erg \, s^{-1}$,
  which might be exceeded in the case presented here. It also means
  that the mass estimate of MCG--05--23--016 based on the temporal
  behaviour is at the lower limit of the possible values.

Figure~\ref{fig:SED} shows the
SED of the source for the data presented here, corrected for Galactic
absorption of $N_{H,Gal} = 8 \times 10^{20} \rm \, cm^{-2}$ in the
line of sight, giving an extinction of $A_V = N_{\rm H} / 1.79 \times
10^{21} \rm \, cm^{-2} = 0.45 \rm \, mag$ (\cite{extinction}). The
extinction in the UV is higher (e.g. $A_U = 0.74 \rm \, mag$), while the effect is insignificant in
the near infrared with e.g. $A_K = 0.05 \rm \, mag$ (Schlegel, Finkbeiner, \& Davis 1998).
To the simultaneous data we added previous observations
in the J, H, and Ks band from the 2MASS and VLA observations at 6 cm
and 20 cm (\cite{Ulvestad84}) for comparison. The SED indicates that the
bolometric luminosity is probably $L_{bol} > 2 L_X$. Even
considering the uncertainty of the mass determination, the source cannot be
assumed to be a typical low-luminosity AGN, which exhibits Eddington
ratios in the range $10^{-3}$ to $10^{-6}$ (Ho 1999), and the Seyfert
1.9 shows an Eddington ratio $10^8$ times larger than that of \object{Sgr A*}
which has a similar mass of $M = 3.3 \times 10^6 \rm \, M_\odot$
(Sch\"odel et al. 2002). If we assume that the mass is in fact $M
  \lae 5 \times 10^6 \rm \, M_\odot$ then the Eddington ratio of
  $L_{bol}/L_{Edd} \gae 0.8$ is also remarkable when
compared to other Seyfert galaxies. Woo \& Urry (2002) list a total of
32 Seyfert galaxies with black hole masses lower than $10^7 \, \rm
M_\odot$ and only 3 of them have $L_{bol}/L_{Edd} \gae 1$, whereas
the average of these lower mass black holes is $L_{bol}/L_{Edd} = 0.30 \pm 0.05$.
In the Galactic
equivalent, the hard state is usually reached at small Eddington
ratios, like $L_{bol}/L_{Edd} \simeq 10^{-3}$ in the case of \object{XTE~J1118+480} (Esin et
al. 2001), 
and $0.003 - 0.2$ for \object{XTE J1550+564} (Done \& Gierli\'nski
2003). 
In fact the SED of the X-ray nova XTE~J1118+480 in the low state appears very similar to the one of
MCG--05--23--016. In XTE~J1118+480 the thermal disc component has a
temperature as low as 24 eV. Considering that the temperature scales
with $T \propto M^{-0.25}$, the disc emission of the Seyfert can be expected at much
lower frequencies and can be hidden in the SED component peaking
around $10^{14.5} \rm \, Hz$. XTE~J1118+480 also shows a power law
component dominating the spectrum in the $0.4 - 160 \rm \, keV$ range
with a photon index of $\Gamma = 1.8$ (\cite{McClintock01}). The overall
SED in this object can be explained by an advection dominated
accrection flow (ADAF) model at 2\%
Eddington ratio (\cite{Esin01}). Although the SED of MCG--05--23--016
appears to resemble the one of XTE 1118+480, its high Eddington ratio
is unlikely to be arising from a radiative inefficient accretion as
described in the ADAF. 
An object with higher Eddington ratio also in the hard state is
\object{GX 339-4}, which reaches $L_{bol}/L_{Edd} = 0.25$ in the hard
state in some cases before the state transition into the high-soft
state (Zdziarski et
al. 2004), consistent with MCG--05--23--016, although on average GX 339--4 reaches only an Eddington
ratio of $0.015$ and $0.05$ during quiescence and during outburst,
respectively. 
Considering a time scale for accretion rate changes
in GX 339--4 of $\sim 1000$ days (Zdziarski et
al. 2004), this would correspond to time scales of $\sim 1$ Myr in the
case of MCG--05--23--016, as the variation times scale with the mass of the
black hole.


Objects which are thought to exhibit extremely high Eddington ratios
are the ultra-luminous X-ray sources (ULX). The true nature of these
objects is still unclear though. Considering their luminosity, they
are either intermediate mass black holes (IMBH) with a central mass as
high as $100 - 100,000 \rm \, M_\odot$ (e.g. Colbert \& Mushotzky
1999), or alternatively they are operating at very high Eddington
ratio, as large as   $L_{bol}/L_{Edd} > 20$ (\cite{Soria07,Roberts07}).
ULX also show hard spectra and low temperature disc
similar to XTE~J1550--564 in very high state where the disc
temperature decreases (e.g. Kubota \& Done 2004). A study of {\it
  Chandra} data on ULX showed that the spectra can often be fitted by a
simple power-law model, without evidence for thermal accretion disc
components (\cite{Berghea08}), similar to MCG--05--23--016. In some
cases ULX show a disc component, e.g. the {\it XMM-Newton} observation
of two ULX in \object{NGC~1313} revealed soft components which are
well fitted by multicolor disc blackbody models with color
temperatures of $kT \simeq 150 \rm \, eV$. 

An alternative model for the accretion process onto black holes has
been proposed by Courvoisier \& T\"urler (2005), in which the
accretion flows consist of different elements (clumps) which have
velocities that may differ substantially. As a consequence, collisions
between these clumps will appear when the clumps are close to the
central object, resulting in radiation. In the case of
MCG--05--23--016, this model results in 
low-energetic collisions, which is also indicated by the missing
variability in the UV band as seen by {\it Swift}/UVOT, where these
collisions should cause flux variations.

Low mass AGN like MCG --05--23--016 operating at high Eddington rate
might be an early state in the evolution towards high-mass black holes
as seen in quasars. As the highest measured redshift of a quasar to
date is $z = 6.43$ (Willott et al. 2007), we can assume that these objects with black
hole masses of $M > 10^8 \rm \, M_\odot$ appear in the Universe around
$z \sim 7$. 
If we assume the formation of the
first heavy black holes with $M \sim 10^6 \rm \, M_\odot$ at redshift $z = 10$, we indeed
need high mass accretion rates. An object like MCG --05--23--016, 
  if we consider its mass to be indeed $M \lae 5 \times 10^6 \rm \,
  M_\odot$, with a
constant Eddington ratio of $L_{bol}/L_{Edd} = 0.8$ and starting black
hole mass of $M(z=10) = 10^6 \rm \, M_\odot$ would reach a mass of $M(z=7) = 4 \times
10^8 \rm \, M_\odot$. However, this would require not only the
existence of a super massive black hole with $M \sim 10^6 \rm \,
M_\odot$ at redshift $z = 10$, but also a high
accretion rate over a time span of $3 \times 10^8$ yrs. But even at a
duty cycle of only $20\%$ for AGN activity at $z \ge 7$, objects like
MCG --05--23--016 can evolve to $10^8 \rm \, M_\odot$, and it has to be
taken into
account that the duty cycle of AGN is likely to be larger in the
high-redshift Universe (Wang et al. 2008). On the other hand, it has
to be considered that the environment in which the AGN grows at
redshifts $z>7$ might be significantly different than the one we
observe MCG --05--23--016 at in the local Universe.

\section{Conclusions}

The Seyfert 1.9 galaxy MCG--05--23--016 shows evidence for a variable
reflection component, variable plasma temperature and high-energy
cut-off energy. The combined data set presented here shows a low reflection
component ($R < 0.3$) compared to previous studies which show a value
as high as $R = 1$. The data require a high-energy cut-off at 72 keV,
but one observation shows an undisturbed power-law up to $E > 100 \rm
\, keV$. Tighter constraints on the hard X-ray spectrum and the
  effect of Compton reflection will be only achievable with focusing
  optics, as they will be provided by the future missions {\it Simbol-X},
  {\it NuSTAR}, and {\it NEXT}.

During the observations the over all luminosity of the hard
X-ray emission remained rather constant, as did the UV/optical
emission. 
The spectral energy distribution shows similarities with the one of
the X-ray nova XTE~J1118+480 in the low-state with the peaks of the
SED shifted according to what one would expect from a scaling of $T
\propto M^{-0.25}$. While XTE~J1118+480 is assumed to accrete at low
efficiency, perhaps through an advection dominated accretion flow, the
Seyfert core is likely to be in a radiative efficient state.
The low mass of the central black hole implies that the accretion onto
the central engine takes place at a high Eddington ratio. An
equivalent in the Galactic black hole class appears to be GX 339--4 in
hard state before state transition,
although the efficiency of the accretion in MCG--05--23--016 is still
higher. In this respect this source shows similarities to ULX. 
These high accretion flows are a key
element in order to understand the rapid growth of black holes in the
early Universe. Objects like MCG--05--23--016 might indicate the early
stages of super massive black holes, in which a strong accretion flow
feeds the central engine. With an Eddington ratio of $\gae 0.8$,
objects like the one presented here can develop into quasars with
$10^8 \rm \, M_\odot$, even considering a duty cycle of only $20\%$,
within a time span of $3 \times 10^8$ yrs. 

\begin{acknowledgements}
We thank the referee Chris Done who gave valuable advice which helped
us to improve this paper.
This research has made use of the NASA/IPAC Extragalactic Database
(NED) which is operated by the Jet Propulsion Laboratory.
\end{acknowledgements}

\end{document}